\documentclass[pre,amsmath,amssymb,aps,reprint,groupedaddress,nofootinbib]{revtex4-2} 
\pdfoutput=1
\usepackage[T1]{fontenc}
\usepackage[utf8]{inputenc}
\usepackage{lmodern}
\usepackage{pgf}
\usepackage{tikz}
\usepackage{scalerel}
\DeclareUnicodeCharacter{2212}{\ensuremath{-}}

\usetikzlibrary{svg.path}
\definecolor{orcidlogocol}{HTML}{A6CE39}
\tikzset{
  orcidlogo/.pic={
    \fill[orcidlogocol] svg{M256,128c0,70.7-57.3,128-128,128C57.3,256,0,198.7,0,128C0,57.3,57.3,0,128,0C198.7,0,256,57.3,256,128z}; 
    \fill[white] svg{M86.3,186.2H70.9V79.1h15.4v48.4V186.2z}
                 svg{M108.9,79.1h41.6c39.6,0,57,28.3,57,53.6c0,27.5-21.5,53.6-56.8,53.6h-41.8V79.1z M124.3,172.4h24.5c34.9,0,42.9-26.5,42.9-39.7c0-21.5-13.7-39.7-43.7-39.7h-23.7V172.4z} 
svg{M88.7,56.8c0,5.5-4.5,10.1-10.1,10.1c-5.6,0-10.1-4.6-10.1-10.1c0-5.6,4.5-10.1,10.1-10.1C84.2,46.7,88.7,51.3,88.7,56.8z}; 
  }
}

\newcommand\orcidicon[1]{\href{https://orcid.org/#1}{\mbox{\scalerel*{ \begin{tikzpicture}[yscale=-1,transform shape] \pic{orcidlogo}; \end{tikzpicture} }{D}}}}

\usepackage[citecolor=blue, colorlinks=true, pdfusetitle, pdfauthor={Vishnu V. Krishnan, Kabir Ramola, Smarajit Karmakar}, pdfsubject={Statistical Physics of Disordered systems}, pdfkeywords={Glass, Shear, Oscillatory, Hessian, Eigenvalue}]{hyperref}

\begin{document}

\title{Annealing effects of multidirectional oscillatory shear in model glass formers}

\author{Vishnu \surname{V. Krishnan}\thinspace\orcidicon{0000-0003-3889-3214}} 
\email{vishnuvk@tifrh.res.in}
\author{Kabir Ramola\thinspace\orcidicon{0000-0003-2299-6219}} 
\email{kramola@tifrh.res.in}
\author{Smarajit Karmakar\thinspace\orcidicon{0000-0002-5653-6328}} 
\email{smarajit@tifrh.res.in}
\affiliation{Centre for Interdisciplinary Sciences, Tata Institute of Fundamental Research, Hyderabad 500046, India}

\date{\today}

\begin{abstract}
    We study the effects of cyclic, athermal quasi-static shear on a model glass-forming system in three dimensions. We utilize the three available orthogonal shear planes, namely $XY, YZ \text{ and } XZ$ to better explore the energy landscape. Using measurements of the stroboscopic $(\gamma = 0)$ energy, we study the effects of using an orthogonal shear direction to perturb unidirectional steady-states. We find that that each sequence of the unidirectional protocol leads to compaction with the universal, $\Delta E \sim N^{-1}$ behavior as a function of the number of cycles, $N$. Additionally we find that cyclic shear utilizing multiple shear planes presents hierarchical compaction, producing progressively lower steady state energies compared to a protocol involving unidirectional cyclic shear alone. Furthermore, with the periodicity of the stroboscopic energy as reference, we show that it is possible to achieve steady state limit-cycles of tunable periodicities using different combinations of the three orthogonal strain directions. We find that such protocols exhibit better annealing as compared to protocols with steady states created using unidirectional shear. Importantly, we find a non-trivial trend in the annealing energy and the period of the steady-state limit-cycle, with an aperiodic protocol appearing to produce the most well annealed states. Finally, we compare the phase diagram of the average steady state energy $\langle E_{\text{S.S.}} \rangle$, as a function of the shearing amplitude  $\gamma_{\max}$, using unidirectional and multidirectional protocols, and find that the universal features are preserved.
\end{abstract}


\maketitle

\section{Introduction}
The energy landscape of glasses exhibits a complex and intricate structure~\cite{sastry1998signatures, debenedetti2001supercooled}. Understanding this landscape arising due to the pairwise interactions between a dense network of particles is crucial to gaining insight into the mechanical and dynamical properties of amorphous materials~\cite{sciortino2005potential, heuer2008exploring}. The nature of the energy landscape plays a crucial role in determining the dynamical properties of glasses, and several hypotheses suggest a fractal structure that can lead to non-trivial phases such as Gardner phases that arise due to the hierarchical distribution of energy minima~\cite{charbonneau2014fractal}. Many hypotheses including replica-symmetry-breaking, suggest features where the curvature of the minimum is positively correlated with its depth~\cite{biroli2009random}. Mode-coupling theories present yet another picture of the landscape, predicting a critical temperature marking a transition between regimes of densely and sparsely populated minima~\cite{das2004mode}. The incredibly high-dimensional nature of the energy landscape of glasses, however, makes numerical as well as theoretical characterizations, challenging.

Molecular dynamics simulations under an isothermal ensemble have long been used to ascertain the properties of glasses. Low temperature simulations and other annealing protocols allow the system to settle into lower energy states. However, the time-scales required to sample the deepest minima diverge as the temperature approaches the glass transition temperature. An alternative method of characterizing the landscape is by utilizing athermal ensembles to traverse the potential energy surface. Athermal Quasistatic Shearing (AQS), where the system is always in an energy minimized state traces the surface of the landscape with a changing global, mechanical deformation~\cite{maloney2004universal, maloney2006aqs}. AQS therefore forms a useful method to study the low-temperature properties and response of materials to deformations. Such perturbations induce transitions to more stable local minima via plastic events leading to hysteresis and memory effects, which allow the system to anneal to lower energy states. An important question that remains unanswered is the nature of the transition-states connecting the minima that constitute the energy landscape. A system subjected to such a quasistatic deformation follows minima on the energy landscape, and samples new minima via saddle-node bifurcations that manifest as plastic events~\cite{dasgupta2012universality}. Understanding the nature of such instabilities with varying strain parameters can lead to a direct insight into the connectivity of the energy landscape.

In this paper, we simulate three dimensional amorphous model systems and study the effects of utilizing all the three available shear directions on the behavior of such systems under cyclic shear. Although recent studies have determined the consequences of employing alternating shear orientations at finite temperatures~\cite{priezjev2019accelerated, priezjev2020alternating, das2020annealing}, athermal systems which are always precisely at local energy minima (mechanical equilibrium), remain to be studied. In Section~\ref{sec_uni} we discuss the current understanding of disordered landscapes that emerges from unidirectional cyclic shear simulations. In Section~\ref{sec_pulse} we detail the effects of orthogonally perturbing a system being cyclically sheared along one shear plane. In Section~\ref{sec_steady} we describe the nature of steady states achieved under cyclic shearing protocols involving combinations of shear directions. In Section~\ref{sec_diagram}, we illustrate the effect on the steady state energy - shear amplitude diagram that emerges as a result of using multidirectional protocols.

\begin{figure*}[t!]
    \centering
    \definecolor{ngreen}{rgb}{0.27, 0.75, 0.25}
    \definecolor{nblue}{rgb}{0.21, 0.45, 0.72}
    \definecolor{nred}{rgb}{0.90, 0.10, 0.10}
    \begin{tikzpicture}
        [scale=0.49, Red/.style ={draw=nred, -}, Dashes/.style ={-, dashed}, RDashes/.style ={draw=nred, -, dashed}, BDots/.style ={draw=nblue, -, thick, dotted}, BDDots/.style ={draw=nblue, -, thick, densely dotted}, Axis/.style ={draw={rgb,255: red,128; green,128; blue,128}, thin, ->}, Thin Dashes/.style ={-, draw={rgb,255: red,64; green,64; blue,64}, very thin, loosely dashed}]
        \node (66) at (-5, 6) {\textbf{(iii)}};
        \node (0) at (-3, 4) {};
        \node (1) at (-3, 3) {};
        \node (2) at (-3, 2) {};
        \node (3) at (-3, 1) {};
        \node (4) at (-3, 0) {};
        \node (5) at (-3, -1) {};
        \node (6) at (-3, -2) {};
        \node (7) at (-3, -3) {};
        \node (8) at (2, 0) {};
        \node (9) at (3, -1) {};
        \node (10) at (4, -2) {};
        \node (11) at (5, -3) {};
        \node (12) at (6, 5) {};
        \node (13) at (3, 2.75) {};
        \node (14) at (4, 3.75) {};
        \node (15) at (5, 4.5) {};
        \node (16) at (-3, 6) {};
        \node (17) at (-3, -5) {};
        \node (18) at (-4, -4) {};
        \node (19) at (7, -4) {};
        \node (20) at (1, -4.75) {$\gamma_{\max}$};
        \node (21) at (-4.25, 1) {\large $\langle E_{S.S} \rangle$};
        \node (34) at (2, -4) {};
        \node (35) at (2, -3.75 ) {\scriptsize $\gamma_{\text{yield}}$};
        \node (67) at (-15, 6) {\textbf{(ii)}};
        \node (22) at (-14, 6) {};
        \node (23) at (-14, -5) {};
        \node (24) at (-15, -4) {};
        \node (25) at (-5, -4) {};
        \node (26) at (-15, 1.25) {\large $\langle E \rangle$};
        \node (27) at (-10.75, -4.75) {$\log{N_{\text{cycle}}}$};
        \node (28) at (-11, 1) {};
        \node (30) at (-13, 4.75) {};
        \node (31) at (-11.75, 1.325) {};
        \node (32) at (-14, 1) {};
        \node (33) at (-13, 0.5) {\scriptsize $\langle E_{S.S.} \rangle$};
        \node (57) at (-9.75, -0.75) {};
        \node (58) at (-9, -1) {};
        \node (59) at (-10.825, 0.75) {};
        \node (60) at (-8.775, -1.225) {};
        \node (61) at (-8.25, -1.75) {};
        \node (62) at (-7, -2) {};
        \node (63) at (-6.75, -2.25) {};
        \node (64) at (-6.25, -2.5) {};
        \node (65) at (-6, -2.5) {};
        \node (68) at (-28, 6) {\textbf{(i)}};
        \node (35) at (-27, -4) {};
        \node (36) at (-27, 3) {};
        \node (37) at (-20, -4) {};
        \node (38) at (-20, 3) {};
        \node (39) at (-25.5, 5) {};
        \node (40) at (-18.5, 5) {};
        \node (41) at (-18.5, -2) {};
        \node (42) at (-24.75, 3) {};
        \node (43) at (-23.5, 5) {};
        \node (44) at (-27, -4) {};
        \node (45) at (-18, 3) {};
        \node (46) at (-16.5, 5) {};
        \node (47) at (-26.25, 4) {};
        \node (48) at (-27.75, 2) {};
        \node (49) at (-19.25, 4) {};
        \node (50) at (-20.75, 2) {};
        \node (51) at (-23.5, -4.5) {x};
        \node (52) at (-18.75, -3.25) {z};
        \node (53) at (-26.5, 0.25) {y};
        \node (54) at (-17.75, 4.5) [text=nred] {$XY$};
        \node (55) at (-24.0, 2.5) [text=nblue] {$YZ$};
        \draw [Red, in=90, out=0] (0.center) to (8.center);
        \draw [Red, in=105, out=0] (1.center) to (8.center);
        \draw [Red, in=120, out=0] (2.center) to (8.center);
        \draw [Red, in=150, out=0] (3.center) to (8.center);
        \draw [Red] (4.center) to (8.center);
        \draw [Red] (5.center) to (9.center);
        \draw [Red] (6.center) to (10.center);
        \draw [Red] (7.center) to (11.center);
        \draw [Red, bend left] (8.center) to (12.center);
        \draw [Dashes] (15.center) to (11.center);
        \draw [Dashes] (13.center) to (9.center);
        \draw [Dashes] (14.center) to (10.center);
        \draw [Axis] (17.center) to (16.center);
        \draw [Axis] (18.center) to (19.center);
        \draw [Thin Dashes] (8.center) to (34.center);
        \draw [Red] (30.center) to (31.center);
        \draw [Red, in=-180, out=-60] (31.center) to (28.center);
        \draw [Axis] (23.center) to (22.center);
        \draw [Axis] (24.center) to (25.center);
        \draw [Thin Dashes] (32.center) to (28.center);
        \draw [Red, in=-180, out=-60] (57.center) to (58.center);
        \draw [BDDots] (28.center) to (59.center);
        \draw [Red] (59.center) to (57.center);
        \draw [BDDots] (58.center) to (60.center);
        \draw [Red] (60.center) to (61.center);
        \draw [Red, in=-180, out=-45, looseness=0.75] (61.center) to (62.center);
        \draw [BDDots] (62.center) to (63.center);
        \draw [Red, in=180, out=-45, looseness=0.75] (63.center) to (64.center);
        \draw [Red] (64.center) to (65.center);
        \draw (36.center) to (35.center);
        \draw (35.center) to (37.center);
        \draw (36.center) to (38.center);
        \draw (38.center) to (37.center);
        \draw (36.center) to (39.center);
        \draw (38.center) to (40.center);
        \draw (37.center) to (41.center);
        \draw (40.center) to (41.center);
        \draw (39.center) to (40.center);
        \draw [style=RDashes] (44.center) to (42.center);
        \draw [style=RDashes] (42.center) to (43.center);
        \draw [style=RDashes] (42.center) to (45.center);
        \draw [style=RDashes] (43.center) to (46.center);
        \draw [style=RDashes] (46.center) to (45.center);
        \draw [style=RDashes] (45.center) to (37.center);
        \draw [style=RDashes] (46.center) to (41.center);
        \draw [style=BDots] (47.center) to (48.center);
        \draw [style=BDots] (49.center) to (50.center);
        \draw [style=BDots] (47.center) to (49.center);
        \draw [style=BDots] (48.center) to (50.center);
        \draw [style=BDots] (48.center) to (44.center);
        \draw [style=BDots] (50.center) to (37.center);
        \draw [style=BDots] (49.center) to (41.center);
    \end{tikzpicture}
    \caption{{\bf (i)} Two of the three fundamental shear deformations available to a three dimensional system. We use these deformations to perform mechanical annealing of the system.  {\bf (ii)}  Schematic of the hierarchical compaction in energy achieved using multi-directional cyclic shearing. Applying unidirectional cyclic alone yields a compaction with a change in energy $\Delta E \sim N_{\text{cycle}}^{-1}$ with increasing number of cycles $N$. This behavior saturates after a critical number of cycles beyond which the system does not anneal further. Surprisingly, providing a cyclic perturbation along an orthogonal shear plane releases the system from a limit cycle, and additional compaction occurs along the original direction  with the same  $\Delta E \sim N_{\text{cycle}}^{-1}$  behavior. In this manner, a hierarchy of energy cascades can be achieved. {\bf (iii)} Phase diagram depicting the unidirectional steady state energy $\langle E_{\text{S.S.}} \rangle$ as a function of the strain amplitude $\gamma_{\max}$. For $\gamma_{\max} < \gamma_{\text{yield}}$ the system anneals to a steady state energy, whereas it rejuvenates for larger amplitudes. For average energies less than the critical threshold the system does not display any variation until yield.}\label{fig_scheme}
\end{figure*}
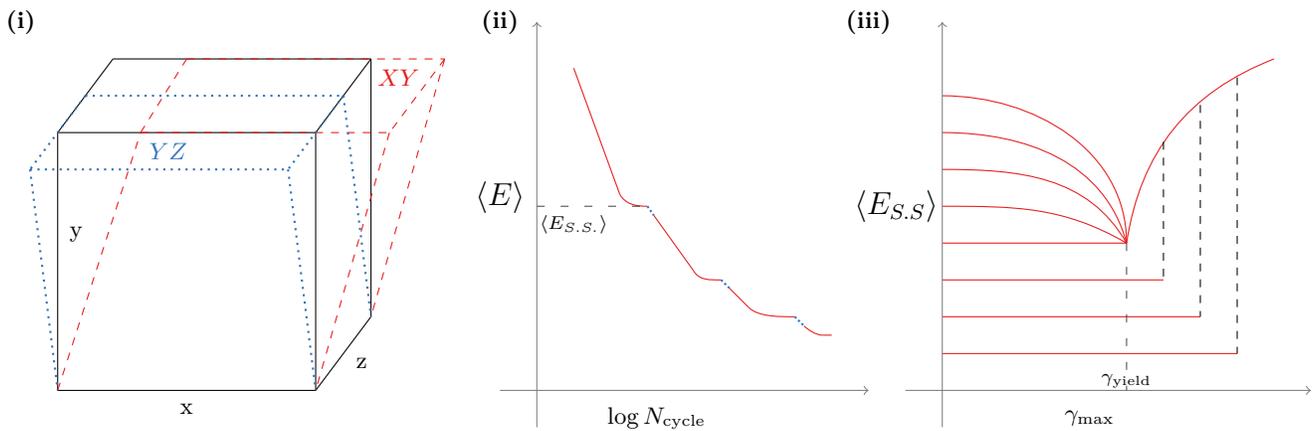

\section{Annealing effects of unidirectional cyclic shear}\label{sec_uni}

A measure of the stability of a system is the capacity of its energy landscape to sustain a perturbation and return to its original state. Probes of the mechanical stability in-turn provide useful information about the landscape, and cyclic AQS forms exactly such a mechanism~\cite{lacks2004energy}. As the dynamics is performed at zero temperature (athermal), the system is always at an energy minimum, and the constraints of mechanical equilibrium are exactly imposed at the microscopic level. Repetitive, oscillatory deformations carry the system through various locations of the complex energy landscape and eventually trap the system into a limit-cycle, which depends on the driving protocol~\cite{regev2015reversibility}. The system settles into a limit cycle that is stable to the corresponding oscillation protocol, with the depth of the stable minima being correlated with the amplitude of the cycle. Several studies of unidirectional oscillatory shear have observed that there exists a critical strain amplitude below which the system settles into states of lower energies (anneal) with increasing number of cycles, and above which, oscillatory shear takes the system through states of higher energies (rejuvenate)~\cite{regev2013onset, fiocco2013oscillatory, keim2014mechanical, regev2015reversibility, lavrentovich2017period, leishangthem2017yielding, bhaumik2021role}. This critical strain $(\gamma_{\text{yield}})$ corresponds to a system-spanning yield event, and recent studies have established that cyclically shearing with an amplitude $\gamma_{\max} < \gamma_{\text{yield}}$ always induces a limit cycle. For a fixed strain orientation, at low amplitudes, the period of the limit cycle is one. However, as the critical strain is approached, the period may increase, along with the time taken to find this steady state. Above the critical strain, the system is chaotic, and does not settle into a limit cycle~\cite{regev2015reversibility,  leishangthem2017yielding}.

In this work, we employ a simple model glass former described in detail in Appendix~\ref{app_model}, in order to conduct simulations of cyclic shear. This model system incurs a system-spanning yield event close to a strain $\gamma_{\text{yield}} = 7 \times 10^{-2}$. 
We begin by examining the evolution of the energy of the system under cyclic shearing at a small strain amplitude, $\gamma_{\max} = 5\times 10^{-2}$, smaller than the yield strain $\gamma_{\text{yield}} \approx 7 \times 10^{-2}$. The energy is measured at $\gamma=0$ after each strain cycle that comprises of a change in strain from $0 \to \gamma_{\max} \to -\gamma_{\max} \to 0$, performed in strain steps of $\Delta \gamma = 5 \times 10^{-5}$, while the energy of the system is minimized with respect to the particle-position degrees of freedom, at every step. During such an oscillation, the system traverses the energy landscape, undergoing elastic as well as plastic deformations. These plastic events allow the system to transition between various local minima and explore a larger region of the landscape as determined by the strain amplitude. The energy of the system at zero strain serves as a measure of the stability of the new minima that are accessed.


Utilizing a single shear plane allows us to reproduce features previously observed in models of amorphous solids, and are described in detail in Secs.~\ref{sec_steady}~and~\ref{sec_diagram}. Cyclically shearing at amplitudes below yield leads the system to find steady state trajectories that are limit-cycles where every configuration settles into a closed trajectory in the state of the system with respect to the driving. Amplitudes larger than the yield strain on the other hand, lead to chaotic trajectories.

\begin{figure*}[t!]
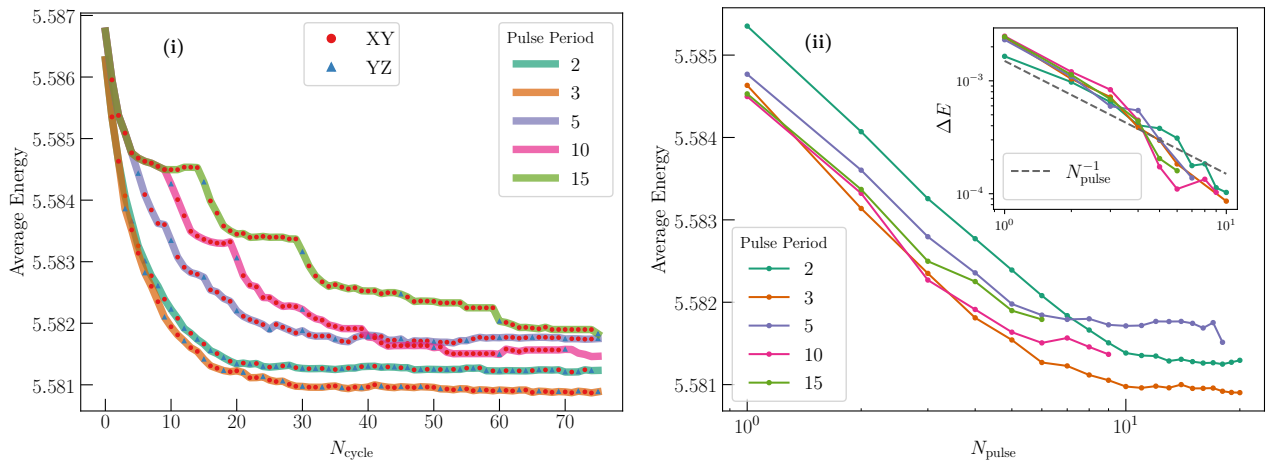

    \centering
    \resizebox{0.47\linewidth}{!}{\input{pulse_annealing_cyclicAQS_3dR10_04096_GMAX5e-2.pgf}}
    \resizebox{0.47\linewidth}{!}{\input{pulses_deltaEnergy_cyclicAQS_3dR10_04096_GMAX5e-2.pgf}}
    \caption{\textbf{(i)} Averaged, stroboscopic energies achieved using a \textit{pulsed} protocol, where unidirectional cyclic shear along $XY$ is punctuated with periodic pulses of cyclic shear along the orthogonal plane $YZ$, both with an amplitude $\gamma_{\max} = 5\times 10^{-2}$. \textbf{(ii)} Energy measured just before a pulse, plotted against the pulse number. \textbf{(Inset)} Change in energy produced by each pulse. The difference in energy between the state just before a pulse and the state just before the next pulse.}\label{fig_pulse}
\end{figure*}

On average, the system anneals when driven with an amplitude lower than $\gamma_{\text{yield}}$, and rejuvenates at larger amplitudes. These features are captured within a steady state phase-diagram of the average steady state energy $\langle E_{\text{S.S.} \rangle}$ as a function of the amplitude of driving $\gamma_{\max}$ (see Fig. \ref{fig_scheme} {\bf (iii)}). While the degree of annealing improves up until the point of yielding, there exists a lowest energy achievable by such mechanical perturbations. In fact, using an initial ensemble with energies lower than this threshold energy leads to no changes to the average energy. Further, the yield strain also increases as the average energy of the ensemble is lowered below this threshold.

Although cyclic shear has now been established as a physically relevant protocol with which to sample the energy landscape and anneal glassy systems, the nature of the energy cascades induced by shearing along multiple planes has received less attention. An interesting aspect of the approach to the steady state in cyclically driven systems is the logarithmic relaxation of the energy, indicating an underlying \textit{compaction} of states. The steady states themselves manifest as a pinch-off from the logarithmic energy decay. This observation is suggestive of an underlying compaction of the final states reached during the process of cyclic shearing, as has also been seen in granular systems subjected to cyclic deformation~\cite{bandi2018training, nicolas2000compaction, pouliquen2003fluctuating}. Indeed, as we show in the next section, the energy landscape sampled by multidirectional shearing presents a hierarchical structure which enables the system to anneal into better minima that unidirectional shear alone cannot access.

\section{Hierarchical Compaction using orthogonal perturbations}\label{sec_pulse}

The behavior of systems under cyclic driving along a single direction have been well characterized, with the system settling into a limit cycle after a large number of oscillations. Limit-cycles trap the system in trajectories that correspond to steady states, with the number of cycles required providing information about \textit{ease} of finding such closed paths. As a three-dimensional solid possesses three orthogonal shear planes, we can perform a more comprehensive exploration of the energy landscape by using various combinations of independent shear cycles along each of the three directions. These additional orthogonal shear directions allow the system to explore new minima of the energy landscape, and therefore studying the effect of perturbing a system being driven under unidirectional cyclic shear provides a further characterization of the nature of the connectivity of the underlying landscape of unstrained states.

To achieve this, we begin with a system corresponding to a parent temperature $T_{P} = 0.58$, undergoing repetitions of cyclic AQS at an amplitude $\gamma_{\max} = 5\times 10^{-2}$ along one shear plane, $XY$. This protocol leads to a logarithmic relaxation of the energy, with a steady state energy being achieved after a finite number of cycles ($\approx 15$ for these parameters). In order to test the stability of this steady state, we periodically apply cycles of the same amplitude after a designated number of cycles along an orthogonal direction, $YZ$. We find that these orthogonal perturbations free the system from the limit cycle, and allow it to perform additional relaxations along the original shear direction. This points to the fact that the steady states achieved by unidirectional shear alone do not fully compactify the system.

We next perform a systematic study of this additional compaction by introducing orthogonal perturbations at specific intervals into a unidirectional shearing protocol, which we term as `pulses'. A pulse-period of $2$ is where every second cycle is a pulse, a pulse-period of $3$ where every third cycle is a pulse, and so on. In Fig.~\ref{fig_pulse}~\textbf{(i)} we plot the energy at zero strain, after each cycle, for protocols with varying pulse periods. At this amplitude of oscillation and the ensemble of energies chosen, all protocols display an annealing of the energy. We find clear signatures of two regimes: (i) at small periods, where the system is unable to anneal completely with unidirectional shear and therefore the steady state limit-cycles depend on the precise protocol as seen in the case of pulse-periods $2, 3$, and (ii) at large periods where the system finds limit-cycles corresponding to the unidirectional protocol, before the perturbation unsettles it, as seen with pulse-periods $10, 15$. There additionally exists an intermediate, transitory regime (such as pulse-period $5$) where the reducing relaxation times cross-over from a value greater than the pulse-period, to a lesser value, as the driving progresses.

Surprisingly, within the number of cycles probed, the degree of relaxation achieved by the unidirectional cyclic protocol depends non-monotonically on the period of the perturbation. We find that, in ascending order of steady-state energy, the periods are: 3, 2, 10, 5, 15. The rates of relaxation though, vary as expected, with the protocols of higher pulse-periods taking longer to reach their steady states. This occurs since the system is able to find a novel steady state where the unidirectional limit-cycle is stable to orthogonal perturbations. Such a state is distinct from multi-directional limit-cycles, and can possibly explain the aforementioned, non-monotonic annealing behavior. We also note that when the pulse periods are larger than the relaxation times, all the trajectories are identical, and may be scaled by the period.

In Fig.~\ref{fig_pulse}~\textbf{(ii)}, we plot the average energy that the system is allowed to relax to, via unidirectional cyclic shear, as measured just before a pulse. The system exhibits an initial, logarithmic annealing in the energy, as has also been observed in finite temperature systems~\cite{das2020annealing, priezjev2021yielding}. In the inset, we plot the change in energy after each pulse, which displays a scaling consistent with compaction: $\Delta E \sim N^{-1}_{pulse}$. As the change in energy due to the unidirectional shearing between each pulse also displays such a compaction, we conclude that the addition of orthogonal shear planes creates a hierarchy of compaction, with the total change in energy achieved by unidirectional shear decreasing with each additional pulse. The orthogonal perturbations therefore increase the number of unidirectional cycles required to achieve a steady state, extending the annealing regime, allowing the system to find lower energy states.


While the protocols studied in this Section considers shearing along orthogonal directions as perturbing pulses, it is natural to also consider finite sequences of orthogonal shearing that can allow the system to explore the full hierarchy of compaction. For example a sequence of $XY$ cycles interspersed with $YZ$ cycles, may further be interspersed with $XZ$ cycles, and can lead to even lower energy states. Such protocols can also lead to non-trivial steady states that are not achievable through unidierctional shearing alone, as we discuss below.

\section{Steady states achieved using multidirectional protocols}\label{sec_steady}

\begin{figure*}[t!]
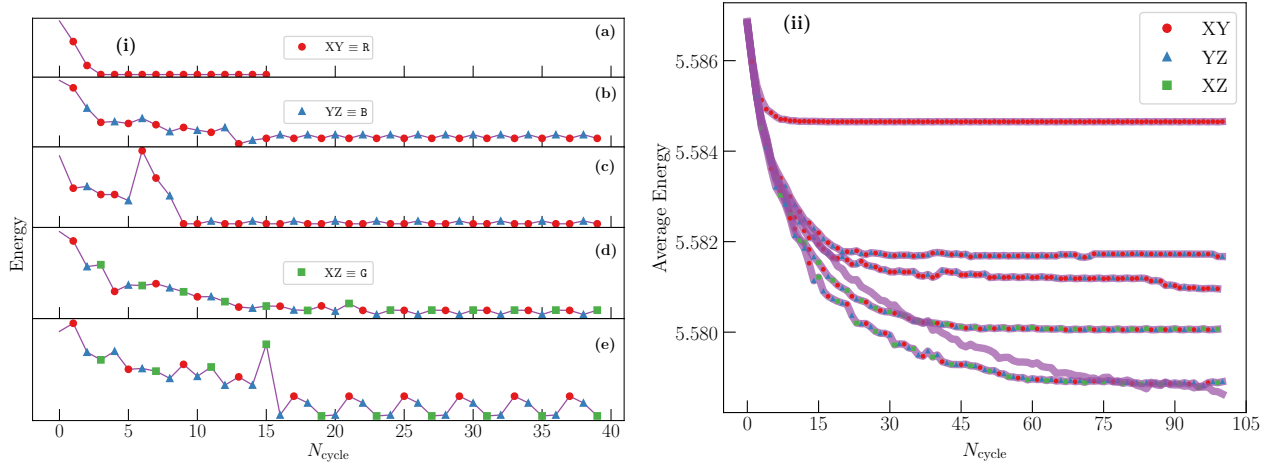

    \centering
    \resizebox{0.47\linewidth}{!}{\input{periods_cyclicAQS_3dR10_04096_GMAX5e-2.pgf}}
    \resizebox{0.47\linewidth}{!}{\input{annealing_cyclicAQS_3dR10_04096_GMAX5e-2.pgf}}
    \caption{\textbf{(i)} Typical trajectories of the stroboscopic $(\gamma=0)$ energy during a cyclic, athermal quasi-static shearing (AQS) protocol of amplitude $\gamma_{\max} = 5 \times 10^{-2}$. The initial configurations correspond to the \textit{R10} model, and were sampled from a thermal ensemble at $T_{P}=0.58$. The oscillatory protocols involve repetitions of cyclic shear along: \textbf{(a)} \texttt{R} \textbf{(b)} \texttt{R-B} \textbf{(c)} \texttt{R-B-R} \textbf{(d)} \texttt{R-B-G} and \textbf{(e)} \texttt{R-B-G-B}. \textbf{(ii)} Averaged trajectories of the stroboscopic energy measurement corresponding to the protocols displayed in \textbf{(i)}. The higher-period protocols show larger degrees of annealing, achieving lower energy steady states.}\label{fig_limit_cycles}
\end{figure*}

Having characterized the effects of orthogonal perturbations on the annealing behavior of unidirectional protocols, we next examine the nature of the steady states that develop as the system is driven with a combination of perturbations along the different shear planes. 
In Fig.~\ref{fig_limit_cycles}~\textbf{(i)}, we display the energy achieved by the system, at the conclusion of each cycle. The three colors and symbols correspond to the three possible shear planes along which the system can be sheared: $XY, YZ \text{ and } XZ$, which we denote by \texttt{R, B \textrm{ and } G} respectively. Subfig.~\ref{fig_limit_cycles}~\textbf{(i)}~\textbf{(a)} shows the energy after each cycle, of a conventional unidirectional cyclic shearing protocol involving repetitions of \texttt{R}, displaying annealing and the approach to a steady state, which represents the system settling into a limit-cycle. 
We next perform repetitions of \texttt{R-B}, alternating between the two directions, as shown in Subfig.~\textbf{(b)}. Such a protocol leads to a steady state limit-cycle of period `two', with the system finding two different minima, say $M_{1} \text{ and } M_{2}$, such that $M_{1} \xrightarrow{\texttt{R}} M_{2}, \text{ and } M_{2} \xrightarrow{\texttt{B}} M_{1}$. Similarly, in Subfigs.~\ref{fig_limit_cycles}~\textbf{(c)}, \textbf{(d)} and \textbf{(e)} we show the results of using driving protocols of higher periods, for instance \texttt{R-G-B, R-G-B-G}, and so on. Most significantly, the existence of steady state limit-cycles of large periods, suggests the existence of a complex network of states. 

While it is possible to construct driving sequences of arbitrary period using multiple shearing directions, the period associated with the steady states achieved by the system are not always commensurate with the period of the driving protocol. This suggests that when using protocols of higher periods, the system may not be sampling as many distinct energy states as the period. 
However, we find that for small periods, this does seem to be the case. We explore the nature of these steady states by driving the system with the simplest sequences composed of the three elementary operations $XY$, $YZ$ and $XZ$, which allow the system to find steady states with much larger periods that any two directions alone. Additionally, it is also interesting to note the fact that high-period protocols display an increase in the energy at short-times, contrary to the behavior of the unidirectional protocol that approaches steady states monotonically.

In Fig.~\ref{fig_limit_cycles}~\textbf{(ii)}, we plot the energy of the system driven under the same protocols illustrated in Fig.~\ref{fig_limit_cycles}~\textbf{(i)}, measured stroboscopically, at $\gamma=0$, over a $100$ cycles, and averaged over $64$ initial configurations. The protocols involving multiple shear directions and higher periods display larger degrees of annealing. Although the approach to the steady state is slower, the final states achieved have much lower energies in comparison to the unidirectional shearing protocol. This enhancement in annealing occurs due to the fact that the system settles into a limit cycle that involves multiple sets of states that the cyclic driving along multiple shear directions is able to access. Additionally, the period of driving with multiple directions biases the system to find limit cycles of corresponding lengths.

As the degree of annealing is correlated with the time taken to reach a limit cycle, which in turn is correlated with the period of driving, this opens up an interesting possibility: does an aperiodic sequence of driving anneal better than periodic sequences? In order to answer this question we also drive the system with a \textit{random} protocol, where, the direction of cyclic shear is randomly chosen from amongst the three possibilities: $\texttt{\{R, B, G\}}$. Remarkably, we see that such a protocol is indeed able to anneal better than the periodic protocols, and it does not reach a steady state within $100$ cycles, as shown in Fig.~\ref{fig_limit_cycles}~\textbf{(ii)}. Of crucial interest is the fact that this protocol is slower to anneal to a given energy, than a suitably chosen, periodic protocol, but is certainly the most unconstrained mechanical protocol available, to anneal an amorphous solid. It would be interesting to explore the effects of such random protocols further, and possibly determine the best possible aperiodic protocol with which to anneal such systems.

\begin{figure*}[t!]
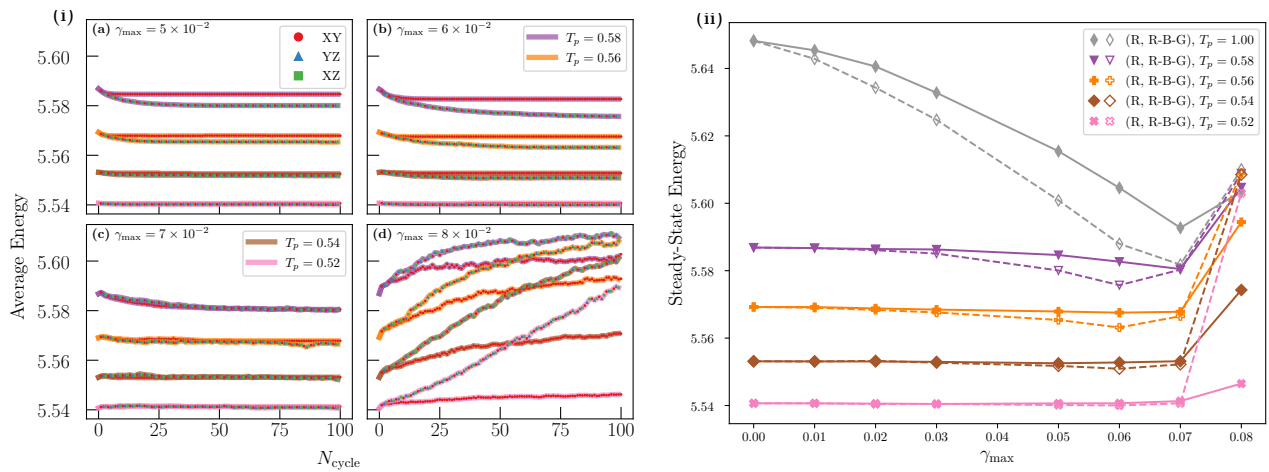

    \centering
    \resizebox{0.482\linewidth}{!}{\input{annealing_cyclicAQS_3dR10_04096_GMAX_TEMP_PERIOD.pgf}}
    \resizebox{0.458\linewidth}{!}{\input{annealing_cyclicAQS_3dR10_04096_GMAX_TEMP_PERIOD_SteadyState.pgf}}
    \caption{\textbf{(i)} Average stroboscopic energy plotted as a function of the number of shear cycles. The initial samples are obtained from various parent temperatures $(T_{p})$, and correspond to the different initial average energies. Two different cyclic protocols (\texttt{R, R-B-G}) and four straining amplitudes $(\gamma_{\max})$ are used. \textbf{(ii)} Steady-state energies plotted against the straining amplitude. Filled markers and solid lines correspond to the \texttt{R} protocol, while unfilled markers and dashed lines correspond to \texttt{R-B-G}.}\label{fig_diagram}
\end{figure*}

\section{Multidirectional annealing phase diagram}\label{sec_diagram}

We have shown in the preceding Sections that utilizing additional shearing directions allows for much better annealing of the model system, at a specific shearing amplitude $(\gamma_{\max} = 5 \times 10^{-2})$, in comparison to unidirectional shearing alone. 
As multidirectional cyclic shear constitutes a more extensive exploration of the potential energy landscape, it is natural to study the effects of such protocols on the nature of the phase diagram of cyclic shear.
In order to identify departures from the $(\langle E_{\text{S.S.}} \rangle, \gamma_{\max})$ phase diagram of steady state energies~\cite{yeh2020glass, bhaumik2021role, sastry2021models}, we examine the behavior of systems cyclically driven under various amplitudes, as well as parent-temperature ensembles.

In Fig.~\ref{fig_diagram}~\textbf{(i)}, we plot stroboscopic energies as a function of the cycle number, comparing two different cyclic protocols: \texttt{R} and \texttt{R-B-G}. We use initial ensembles sampled from the parent temperatures $T_{p} = 0.58, 0.56, 0.54 \text{ and } 0.52$. In Subfig.~\ref{fig_diagram}~\textbf{(i)}~\textbf{(a)} we plot the energy achieved with the shearing amplitude, $\gamma_{\max} = 5 \times 10^{-2}$, also employed in Secs.~\ref{sec_pulse}~and~\ref{sec_steady}. Under unidirectional cyclic shear, lower parent temperatures have been shown to be accompanied by a reduction in the degree of annealing, with a marked threshold energy below which mechanical protocols fail to produce any change in the average energy~\cite{yeh2020glass, bhaumik2021role}. We observe a similar trend when using a multidirectional protocol, with the system presenting an energy below which there is no annealing. In Subfig.~\ref{fig_diagram}~\textbf{(i)}~\textbf{(b)} we display the behavior of the system at a larger shearing amplitude, $\gamma_{\max} = 6 \times 10^{-2}$. Although there are marked enhancements in annealing, the threshold in energy persists. Interestingly, the threshold for the multidirectional protocol coincides with that for the unidirectional protocol. The robustness of the phase diagram is additionally corroborated by the difference between the steady state energies attained by the two protocols decreasing with lowering temperatures. The two protocols yield the same degree of annealing at the threshold energy (Appendix Fig.~\ref{fig_diff}), pointing to the robustness of the threshold energy to the driving protocol.

In Fig.~\ref{fig_diagram}~\textbf{(ii)} we plot the steady state energies achieved by cyclic shearing at various amplitudes, beginning from various initial parent temperature ensembles. We find that the main features of the phase diagram for unidirectional shearing are preserved. However multidirectional shearing displays a significant enhancement in annealing, when employing shear amplitudes below the yield strain $(\gamma_{\text{yield}} \approx 7.0 \times 10^{-2})$. Additionally, configurations sampled at high temperatures show significant annealing that improves as the yield strain is approached. The lowest energy that the cyclic shearing procedure converges to represents the threshold energy below which athermal protocols fail to anneal. Unidirectional protocols show a sharp approach to this threshold, with only amplitudes very close to the yield strain able to reach it. Multidirectional protocols on the other hand present a more gradual approach to this threshold, allowing for a better estimation of both the threshold energy as well as the yield strain.
As the time taken to reach the steady state diverges as the yield strain is approached~\cite{regev2013onset, regev2015reversibility}, it would be interesting to study such a divergence with multidirectional protocols that may present varying degrees of divergence. This could open up the possibility of an optimal choice of protocol for the purposes of annealing.

Finally we examine the rejuvenation effects displayed by the system for amplitudes larger than the yield strain.
In Subfig.~\ref{fig_diagram}~\textbf{(i)}~\textbf{(d)}, we plot the energies achieved when using an amplitude above the yield strain. For the unidirectional protocol, we find an increase in energy consistent with earlier results. In comparison, the multidirectional protocol displays a significant enhancement in rejuvenation of the energy. We note that in the case of the multidirectional protocol, the amplitude at which the system begins to rejuvenate is much lower. The corresponding regime in Fig~\ref{fig_diagram}~\textbf{(ii)} illustrates the difference between the two protocols. In Subfig.~\ref{fig_diagram}~\textbf{(i)}~\textbf{(c)} we plot the intermediate regime between the annealing and rejuvenating phases. As this regime is very close to the yield strain, there is no significant change in the average energy achieved using both protocols. The consistency of this behavior across protocols reaffirms the significance of the yield strain as an inherent property of the model system.


\section{Conclusion and Discussion}

In this paper, we have described the effects of incorporating orthogonal shear planes into athermal cyclic shearing simulations. We have shown that incorporating multiple shear directions into the mechanical annealing of such systems can lead to significant increase in the annealing effects of such glass formers. Our results therefore provide further insight into the nature of states accessed by cyclic shearing. 
Additionally, we uncovered an interesting hierarchy of annealing that such multidirectional protocols can access. We illustrated this by perturbing steady states achieved by unidirectional cyclic shear, which results in an escape from the limit cycle, and leads to further annealing along the original direction. 
We also highlighted the fact that using a protocol with no periodicity, i.e. a random choice of shear plane for every oscillation, exhibits a significant increase in annealing over periodic protocols. This opens up the possibility of designing protocols with arbitrary periods to their steady state limit cycles. The longer times required to find trajectories that are stable to these protocols allows the system to anneal to deeper minima.
Finally, we also illustrated the advantages of using multidirectional protocols in examining the steady state energy phase diagram, and also in estimating the threshold energy. As the singular behavior at the yield strain renders the approach to it difficult, driving the system in multiple directions can lead to a smoother approach to the yield point. As the yield strain sets a bound on the spatial extent of a unidirectional limit cycle, the addition of multiple shear directions allows for the possibility of creating arbitrarily large limit cycles, which in turn can provide greater insight into the nature of yielding in such systems.

An important question that remains to be addressed is the origin of the yield strain and the threshold energy. It would be intriguing to find the relationship between them, and crucially, also with the microscopic parameters of the system such as measures of disorder such as the particle-size polydispersity. Another relevant direction of research is the nature of the connectivity between energy minima accessible via mechanical annealing.
The results from the periodic perturbations suggests the existence of non-trivial, optimal perturbation frequencies that provide the best annealing. It would be interesting to study the nature of the effects of such periodicity in the driving protocol further.
Our study also raises interesting questions regarding the entropy of the limit cycles that are accessed by periodic and aperiodic protocols, which can in turn provide crucial information about the connectivity of the energy landscape of such amorphous systems. Finally, it would be interesting to study the effects of multiple shearing directions in encoding memory~\cite{memory2019keim} in such systems, which could lead to a better understanding of the structural correlations encoded by cyclic shear.

\begin{acknowledgments}
    We thank Srikanth Sastry\thinspace\orcidicon{0000-0001-7399-1835}, Anoop Mutneja\thinspace\orcidicon{0000-0002-5632-3867} and Bhanu Prasad Bhowmik\thinspace\orcidicon{0000-0003-3298-2099} for useful discussions. V.V.K. thanks the Council of Scientific and Industrial Research, India for support via the Shyama Prasad Mukherjee Fellowship (SPM-07/1142(0228)/2015-EMR-1). S.K. would like to acknowledge the support from Swarna Jayanti Fellowship Grants No. DST/SJF/PSA-01/2018-19 and No. SB/SFJ/2019-20/05. This project was funded by intramural funds at TIFR Hyderabad from the Department of Atomic Energy, Government of India.
\end{acknowledgments}

\begin{appendix}

\section{Model and Methods}\label{app_model}

We simulate a 50:50 mixture of two particle types A and B. The interaction potentials are cut-off at a distance
\begin{equation}
    r_{c} = 1.385418025 \; \sigma,
\end{equation}
with the three interaction diameters given by
\begin{align}
    \sigma_{A A} &= 1.0, \nonumber \\
    \sigma_{B B} &= 1.4, \nonumber \\
    \sigma_{A B} &= \sqrt{\sigma_{A A}\sigma_{B B}}.
\end{align}
The number density of particles $N/V = \rho = 0.81$. We use a purely repulsive pairwise potential, smoothed to 2 derivatives at cut-off:
\begin{equation}
    \psi(r;\sigma)  = {\left(\frac{\sigma}{r}\right)}^{10} + \sum_{m=0}^{2} c_{2 m} {\left(\frac{r}{\sigma}\right)}^{2 m}
\end{equation}
where the constants $c_{2m}$ are calculated such that
\begin{equation}
    \left. \frac{d^m \psi}{d r^m} \right\rvert_{r_{c}} = 0 \quad \forall \quad 0 \le m \le 2.
    \label{eq_smooth}
\end{equation}

Simulations of glasses along with the energy minimizations were performed using LAMMPS~\cite{thompson2021lammps, lammps}. The stopping criterion for the minimization was the force 2-norm: $\sqrt{\sum_{i=1}^{N} {\lvert F_{i} \rvert}^{2}}$. Analyses were performed with the help of \textit{NumPy}~\cite{2011numpy, harris2020array, numpy} and \textit{SciPy}~\cite{2020SciPy, scipy} libraries. Plotting was performed using \textit{Matplotlib}~\cite{hunter2007matplotlib, matplotlib}.

In our simulations, we use the glass forming model defined above, in three-dimensions, with $N=4096$ particles, and begin with equilibrating at $T=1.00$. We then cool samples at a slow rate of $\dot{T} \approx 10^{-2}$ to and equilibrate at various parent temperatures $(T_{p})$. We subsequently employ the conjugate gradient algorithm to achieve states energy minimized up to a force tolerance of $1.0 \times 10^{-10}$.

\section{Unidirectional and Multidirectional Annealing}

\begin{figure}[ht]
    \centering
    \resizebox{\linewidth}{!}{\input{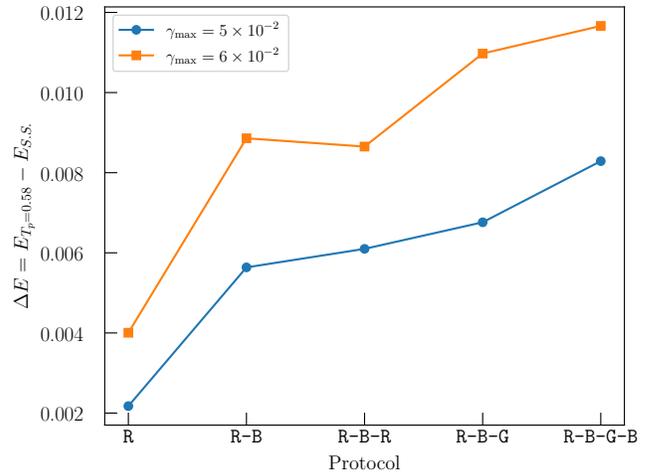}}
    \caption{Degree of annealing plotted as a function of the protocol, for an ensemble sampled from a parent temperature $T_{p} = 0.58$.}\label{fig_proto}
\end{figure}

\section{Unidirectional and Multidirectional Threshold Energies}

\begin{figure}[ht]
    \centering
    \resizebox{\linewidth}{!}{\input{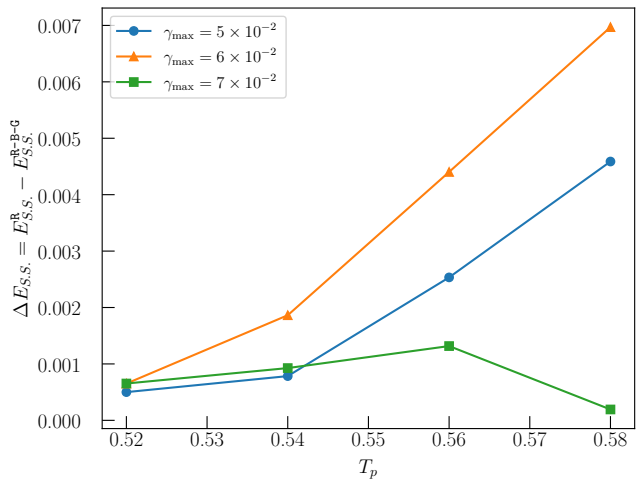}}
    \caption{Difference in annealed energy between the \texttt{R} and \texttt{R-B-G} protocols, plotted as a function of the initial-ensemble temperature, for various oscillation amplitudes.}\label{fig_diff}
\end{figure}

\end{appendix}


\bibliographystyle{apsrev4-2} 
\bibliography{Cyclic3D_Bibliography}

\end{document}